\documentclass[aps,a4paper,twocolumn]{revtex4}
\usepackage{amsmath}
\usepackage{graphicx}
\usepackage{subfigure}
\usepackage[usenames,dvipsnames]{color}
\definecolor{darkblue}{RGB}{0,0,196}
\usepackage[colorlinks=true,linkcolor=darkblue,citecolor=darkblue,urlcolor=darkblue]{hyperref}
\usepackage{setspace}
\usepackage{hyperref}
\usepackage{xcolor}
\hypersetup{
  colorlinks   = true, 
  urlcolor     = red, 
  linkcolor    = blue, 
  citecolor   = blue 
}
\usepackage{graphicx}
\usepackage{amsmath,bbm}
\usepackage{amssymb,bm}
\usepackage{yfonts}
\usepackage{comment}

\begin{document}

\title{Deep learning predicted elliptic flow of identified particles in heavy-ion collisions at the RHIC and LHC energies}
\author{Neelkamal Mallick$^{1,}$\footnote{Neelkamal.Mallick@cern.ch}}
\author{Suraj Prasad$^{1,}$\footnote{Suraj.Prasad@cern.ch}}
\author{Aditya Nath Mishra$^{2,4,}$\footnote{Aditya.Nath.Mishra@cern.ch}}
\author{Raghunath Sahoo$^{1,}$\footnote{Corresponding Author: Raghunath.Sahoo@cern.ch}}
\author{Gergely G\'abor Barnaf\"oldi$^{3,}$\footnote{Barnafoldi.Gergely@wigner.hu}}
\affiliation{$^{1}$Department of Physics, Indian Institute of Technology Indore, Simrol, Indore 453552, India}
\affiliation{$^{2}$Department of Physics, School of Applied Sciences, REVA University, Bangalore 560064, India}
\affiliation{$^{3}$Wigner Research Center for Physics, 29-33 Konkoly-Thege Mikl\'os Str., H-1121 Budapest, Hungary}
\affiliation{$^{4}$Department of Physics, University Centre For Research \& Development (UCRD), Chandigarh University, Mohali, Punjab 140413, India}

\begin{abstract}
Recent developments on a deep learning feed-forward network for estimating elliptic flow ($v_2$) coefficients in heavy-ion collisions have shown the prediction power of this technique. The success of the model is mainly the estimation of $v_2$ from final state particle kinematic information and learning the centrality and the transverse momentum ($p_{\rm T}$) dependence of $v_2$. The deep learning model is trained with Pb-Pb collisions at $\sqrt{s_{\rm NN}} = 5.02$ TeV minimum bias events simulated with a multiphase transport model (AMPT). We extend this work to estimate $v_2$ for light-flavor identified particles such as $\pi^{\pm}$, $\rm K^{\pm}$, and $\rm p+\bar{p}$ in heavy-ion collisions at RHIC and LHC energies. The number of constituent quark scaling is also shown. The evolution of $p_{\rm T}$-crossing point of $v_2(p_{\rm T})$, depicting a change in baryon-meson elliptic flow at intermediate-$p_{\rm T}$, is studied for various collision systems and energies. The model is further evaluated by training it for different $p_{\rm T}$ regions. These results are compared with the available experimental data wherever possible.
\pacs{}
\end{abstract}
\date{\today}
\maketitle 

\section{Introduction}
\label{intro}
Relativistic heavy-ion collisions have been studied extensively for decades in experiments at the Large Hadron Collider (LHC), CERN, Switzerland, and Relativistic Heavy Ion collider (RHIC), BNL, USA. The formation of a deconfined thermalized medium of quarks and gluons has already been established in such collisions~\cite{Bass:1998vz}. This medium of hot and dense state of strongly interacting matter is known as the quark-gluon plasma (QGP). Direct probes for studying the properties of QGP are not available owing to the behavior of the strongly interacting matter. However, the signatures of the formation of QGP could be studied using various indirect effects such as jet quenching, strangeness enhancement, and quarkonia suppression, just to name a few. Another crucial observable, which is widely studied to investigate the properties of QGP in heavy-ion collisions, is the transverse collective flow~\cite{Heinz:2013th}. This transverse collective flow is anisotropic and depends on the equation of state and transport coefficients of the system. Due to the almond-shaped, elliptical nuclear overlap region in noncentral heavy-ion collisions, the initial state has finite spatial anisotropy, further developing the final state momentum anisotropy of the emitted particles. This momentum anisotropy could be expressed as the coefficients of the Fourier expansion of the azimuthal momentum distribution of the produced particles. Finite azimuthal anisotropy has been well observed in heavy-ion collision experiments so far~\cite{STAR:2003wqp,ALICE:2010suc,ALICE:2011ab,ALICE:2014dwt}. The second-order flow coefficient, also known as the elliptic flow ($v_2$), which is believed to be driven mainly by the geometry of the distributed nucleons in the nuclear overlap region during a noncentral heavy-ion collision, has the dominant contribution to the overall azimuthal anisotropy. 

Different hydrodynamic model calculations suggest that elliptic flow builds up in the early partonic phases of the QGP and evolves through hadronic rescatterings in the hadronic phase. Theoretically, azimuthal anisotropy is thus understood to have consequences from (ideal) fluid dynamics applied to the QGP phase and kinetic descriptions applied to the microscopic hadron cascade phase~\cite{Kolb:2000fha,Bass:2000ib,Nonaka:2006yn,Teaney:2001av,Hirano:2005xf}. Therefore, to understand the interplay of partonic and hadronic phases in the evolution of collective flow, it becomes necessary to study the elliptic flow for different identified particles.
The competing effects of radial (symmetric) flow and hadronic rescatterings could also be studied by estimating the elliptic flow of identified particles.

Another exciting thing that has been observed in heavy-ion collisions is that $v_2$ of baryons is larger than that of mesons in the intermediate-$p_{\rm T}$ range ($2.0 \lesssim p_{\rm T} \lesssim 5.0$ GeV/c). This is usually attributed to the particle production by constituent quark coalescence~\cite{Voloshin:2002wa}. According to this theory, the development of hydrodynamic flow happens in the early deconfined partonic phase and is subsequently transferred to the hadrons by hadronization through the quark coalescence mechanism~\cite{Molnar:2003ff,Sato:1981ez,Dover:1991zn}. In an ideal case, this behavior should lead to the number of constituent-quark (NCQ) scaling of the observed flow for baryons and mesons and, thus, strongly hints at the appearance of collectivity at the early deconfined partonic level. In Au-Au collisions at $\sqrt{s_{\rm NN}} = 200$ GeV, the scaling holds only approximately~\cite{PHENIX:2012swz}. However, in Pb-Pb collisions at $\sqrt{s_{\rm NN}} = 2.76$ and $5.02$ TeV, the scaling holds only up to a level of $\pm20\%$~\cite{ALICE:2014wao,ALICE:2010suc,ALICE:2018yph}. 

There are several methods to estimate the flow coefficients, such as the complex reaction plane identification method~\cite{Poskanzer:1998yz}, the cumulant method~\cite{Borghini:2000sa}, the Lee-Yang zeroes method~\cite{Bhalerao:2003xf}, and the principal component analysis method~\cite{Bhalerao:2014mua,Hippert:2019swu,CMS:2017mzx,Gardim:2019iah}. For the first time, it has been shown that deep learning models, based on machine learning (ML) framework, could also be used to estimate the flow coefficients, mainly the results of elliptic flow have been emphasized~\cite{Mallick:2022alr}. 

The motivation of the present study is to prepare a deep learning framework for the estimation of elliptic flow from final state particle kinematic information. This is an attempt to see whether a machine learning model can learn from final state particle correlations to predict any physical observable of interest. Since the elliptic flow is influenced by several factors such as the collision centrality, energy, system size, particle mass, particle species, and transverse momentum \textit{etc.}, it will be interesting to see how good the deep learning model is to preserve these dependencies. The trained model is found to be faster and more efficient in computing the flow coefficient than some of the traditional methods such as the multi-particle cumulant method.

The proposed deep learning model successfully learns and preserves the centrality and transverse momentum ($p_{\rm T}$) dependence of $v_2$. It is also shown to perform steadily and accurately when confronted with a data set with additional noise. Here, in this work, we estimate the elliptic flow coefficient for identified stable light hadrons like $\pi^{\pm}$, $\rm K^{\pm}$, and $\rm p+\bar{p}$ in Au-Au collisions at $\sqrt{s_{\rm NN}} = 200 $ GeV, Pb-Pb collisions at $\sqrt{s_{\rm NN}} = 2.76$, and $5.02$ TeV, and Xe-Xe collisions at $\sqrt{s_{\rm NN}} = 5.44$ TeV for (0-10)\%, (40-50)\%, and (60-70)\% centrality classes by using the deep learning estimator mentioned in Ref.~\cite{Mallick:2022alr}. The NCQ scaling is also studied for these collision systems. The evolution of the $p_{\rm T}$-crossing point of $v_2(p_{\rm T})$, depicting a change in baryon-meson elliptic flow at intermediate-$p_{\rm T}$, is studied for the above-mentioned collision systems at different energies. The prediction capability of the model is further evaluated by training it with particles in different $p_{\rm T}$ regions.

The paper is organized as follows. We briefly introduce the event generation using a multiphase transport model (AMPT) and the target observable, elliptic flow in Section~\ref{sec2}. The proposed deep learning estimator is described in Section~\ref{sec3} along with some performance plots of the model. In Section~\ref{sec4}, we describe the results and conclude with a summary in Section~\ref{sec5}.

\section{Event Generation and target observable}
\label{sec2}
In order to train the deep learning-based ML algorithm to estimate the elliptic flow, we have used a multiphase transport (AMPT) model to simulate the dataset, a short description of which is discussed in this section, along with the elliptic flow.
\subsection{AMPT model}
AMPT is a Monte Carlo-based event simulator that is used to generate ultra-relativistic p-A and A-A collisions at RHIC and LHC energies~\cite{Lin:2004en}. AMPT has four components, namely, initialization of collisions by the heavy-ion jet interaction generator model (HIJING)~\cite{ampthijing}, parton transport by Zhang's Parton Cascade model (ZPC)~\cite{amptzpc}, hadronization of the partons performed by spatial coalescence mechanism in string melting version of AMPT and Lund string fragmentation model in the default version of AMPT~\cite{Lin:2001zk,He:2017tla}, and finally, the hadron transport using a relativistic transport model (ART)~\cite{amptart1,amptart2}. AMPT model is extensively used to study the properties of hot and dense medium formed in relativistic heavy-ion collisions, and the elliptic flow coefficient is one of them. String meting mode of AMPT describes the elliptic flow well in the intermediate transverse momentum region with quark coalescence mechanism for hadronisation~\cite{ampthadron2,ampthadron3,Greco:2003mm}. Hence, in this study, we have incorporated the string melting mode with AMPT version 2.26t9b. AMPT settings used in this work are the same as reported in Refs.~\cite{Tripathy:2018bib,Mallick:2020ium, Mallick:2021wop,Prasad:2022zbr}. We have used the impact parameter slicing definition of the centrality, and the corresponding impact parameter values have been taken from Ref.~\cite{Loizides:2017ack}.

\subsection{Elliptic flow}
Anisotropic flow is one of the key observables in the evolution of QGP medium in noncentral relativistic heavy-ion collisions. The pressure gradient formed in the hot and dense medium due to the initial state spatial anisotropy can transform into final state momentum azimuthal anisotropy. The azimuthal anisotropy depends upon the equation of state as well as the transport coefficients of the medium formed. It can be quantified by the coefficients of Fourier expansion of the azimuthal momentum distribution of particles, given by~\cite{Voloshin:1994mz}: 
\begin{equation}
        \frac{dN}{d\phi}=\frac{1}{2\pi}\Big(1+2\sum^{\infty}_{n=1}v_{n}\cos[n(\phi-\psi_{n})]\Big)
    \label{eqn-invfourierexp}
\end{equation}
Here, $v_{\rm n}=\langle\cos[n(\phi-\psi_{\rm n})]\rangle$ denotes $n ^{\rm th}$ order anisotropic flow coefficient, $\phi$ is the azimuthal angle, and $\psi_{\rm n}$ is the $n ^{\rm th}$ harmonic symmetry plane angle. The second-order anisotropic flow coefficient ($v_{\rm 2}$) is called the elliptic flow and has a major contribution from the initial eccentricity. In order to calculate the elliptic flow event-by-event, we have used the event plane method \cite{Masera:2009zz}. Here, we have set $\psi_{\rm R} = 0$, which makes $v_{\rm 2} = \langle\cos(2\phi)\rangle$. The average is taken over all charged particles in an event.

\begin{figure}[ht!]
\includegraphics[scale=0.45]{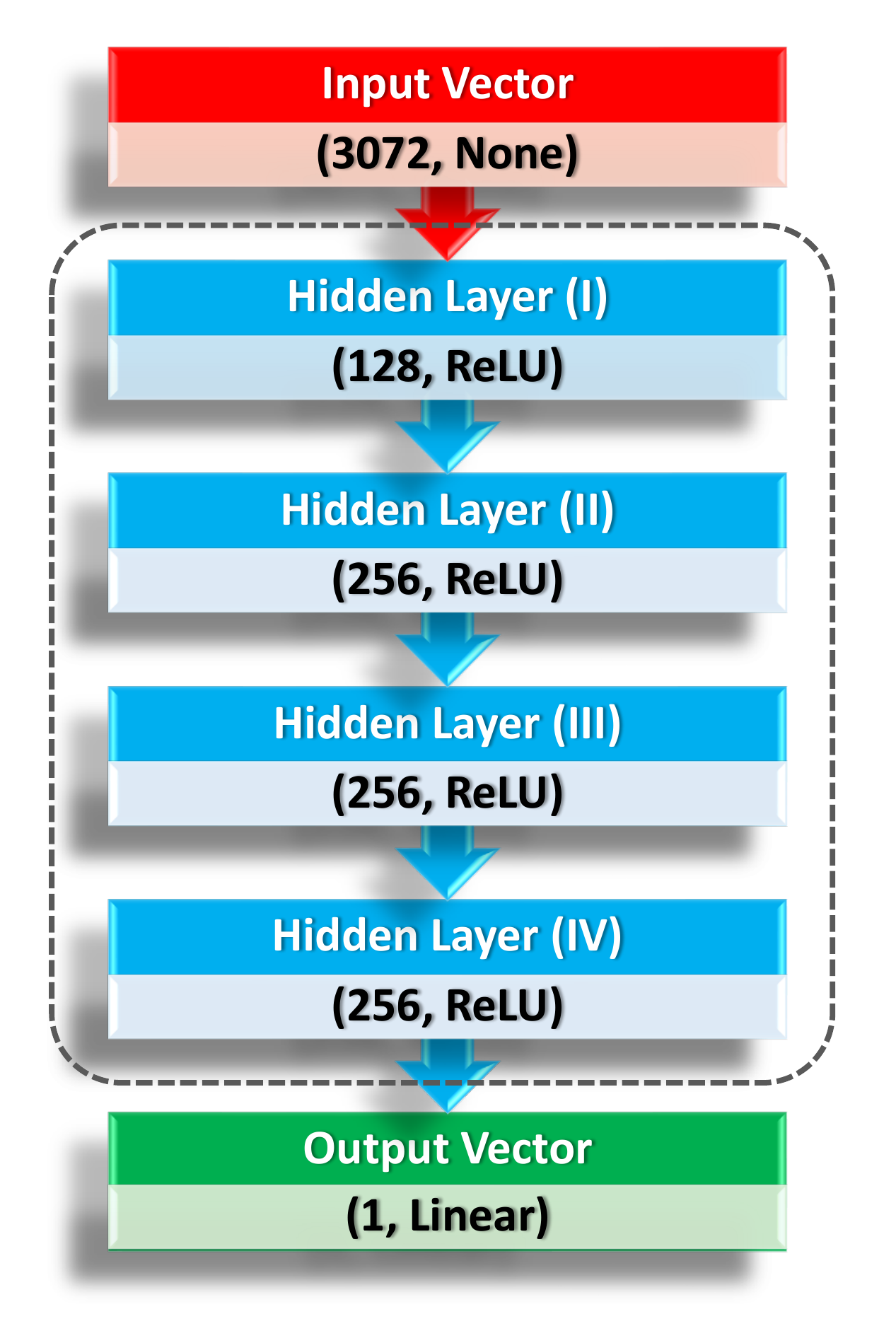}
\caption{(Color online) Schematics of the deep neural network architecture used in this work containing one input layer, followed by four hidden dense layers, and one output layer. The number of nodes and the type of activation function used in each layer is mentioned.}
\label{fig1}
\end{figure}

\section{Deep Learning Estimator}
\label{sec3}

ML is a branch of artificial intelligence that learns correlations from the data to map the input and output observables. Boosted decision trees, deep neural networks, convolutional neural networks, etc., can learn the suitable correlations from the data through model training and help us map complex non-linear observables whose mapping function can not be trivially written in a sequential algorithm. Heavy-ion collisions at LHC energies produce thousands of final state hadrons, each carrying specific kinematic information and evolving through the different physical processes, which make the system complex, non-linear, probing the underlying dynamics challenging. In order to extract the desired physics information from such complex systems, ML has been used by the collider physics community for more than a decade now~\cite{Bhat:2010zz,Roe:2004na,Baldi:2014kfa,Baldi:2014pta,Buckley:2011kc,Bridges:2010de,biro}. In this study, we apply a deep neural network (DNN) based ML algorithm on reckoning the elliptic flow coefficient, which takes the $p_{\rm T}$, mass and $\log\sqrt{(s_{\rm NN}/s_{0})}$ weighted $(\eta-\phi)$ binned values as different features of the input to the model as described in detail in Ref.~\cite{Mallick:2022alr}. The charged particle tracks in $|\eta|<0.8$, with $p_{\rm T} \geq 0.15$ GeV/c and $\phi \in [0, 2\pi]$ are considered for the training. Also, this is an inverse problem of supervised regression kind, thus the output labels of the DNN model, which are the elliptic flow coefficients, are estimated using the event plane method from the AMPT simulation on an event-by-event basis. The model is trained with minimum bias Pb-Pb collisions at $\sqrt{s_{\rm NN}} = 5.02$~TeV. The input data set is divided by the number of events in 8:1:1 for making the training, validation, and testing data set, respectively.

\begin{figure}[ht!]
\includegraphics[scale=0.4]{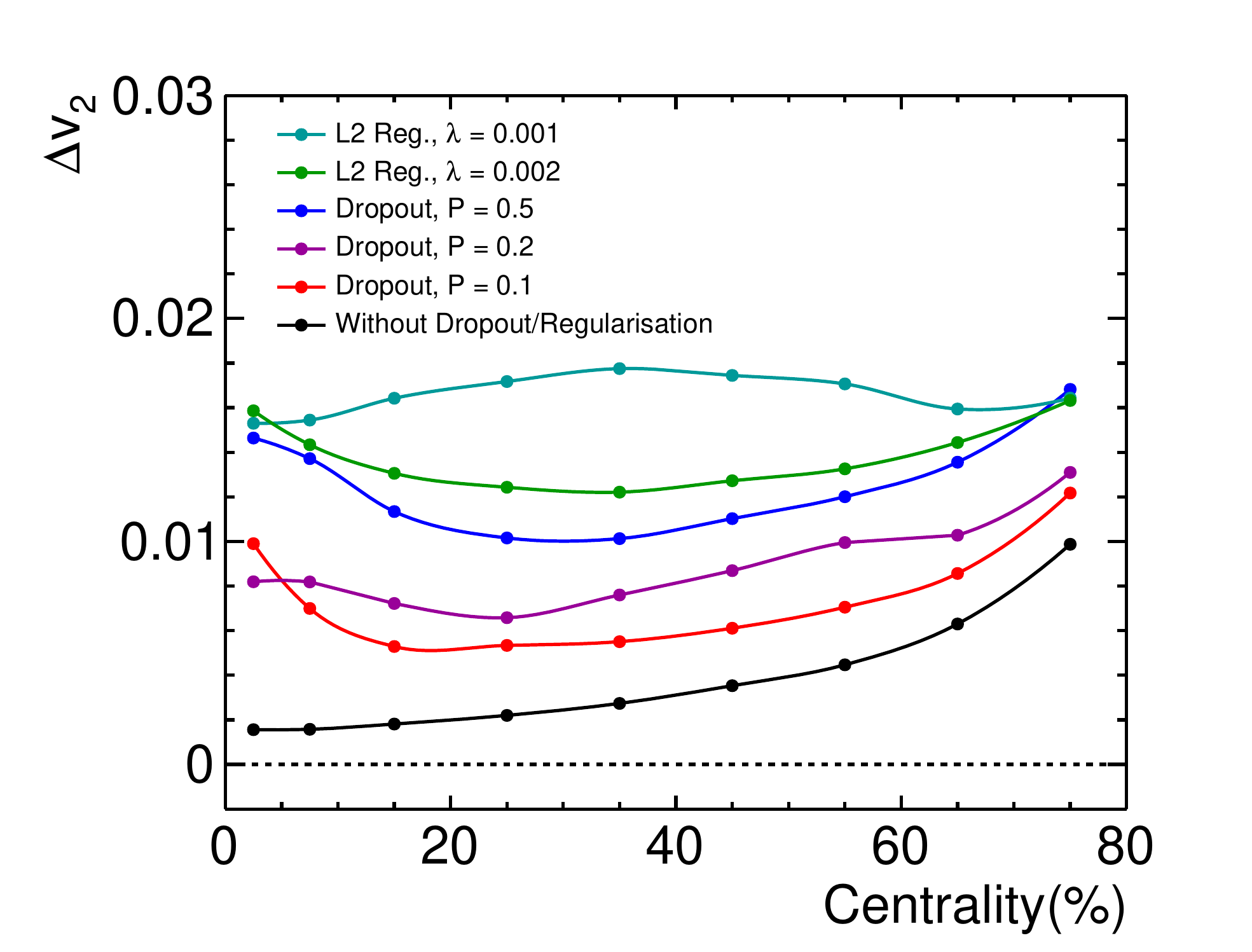}
\caption{(Color online) Evolution of the MAE ($\Delta v_2$) as a function of centrality in Pb-Pb collisions, $\sqrt{s_{\rm NN}} = 5.02$~ TeV with the model trained with L2 regularization, and dropout with their respective hyperparameters.}
\label{fig2}
\end{figure}

The DNN architecture used in this paper for the regression problem consists of one input layer, four dense hidden layers, and one output layer, as shown in Fig.~\ref{fig1}. The input layer with 3072 features are mapped to the output layer via four hidden layers, one after another, each having 128, 256, 256, and 256 nodes, respectively. All the four hidden dense layers use the rectified linear unit as the activation function~\cite{relu}, and the output layer has a single node with a linear activation function. The DNN model uses the \textit{adam} algorithm as the optimizer~\cite{Kingma:2014vow} with mean squared error as the loss function, the details of which can be found in Ref.~\cite{Mallick:2022alr}. We do not implement any L2 regularization or dropout as these seem to hamper the performance of the DNN, as shown in Fig.~\ref{fig2}. The mean absolute error (MAE) of elliptic flow ($\Delta v_2$), defined in Eq.~\ref{eq2}, decreases for an increase in $\lambda$ and decrease in $\rm P$ values; however, $\Delta v_2$ is found to be the smallest for the case with no L2 regularization and dropout. $\lambda$ and $\rm P$ are the hyperparameters in L2 regularization and dropout, respectively.
\begin{eqnarray}
\Delta v_2 = \frac{1}{N_{\rm events}}\sum\limits_{n=1}^{N_{\rm events}}|v_{2_{n}}^{\rm true}-v_{2_{n}}^{\rm pred.}|
\label{eq2}
\end{eqnarray}

\begin{figure}[ht!]
\includegraphics[scale=0.4]{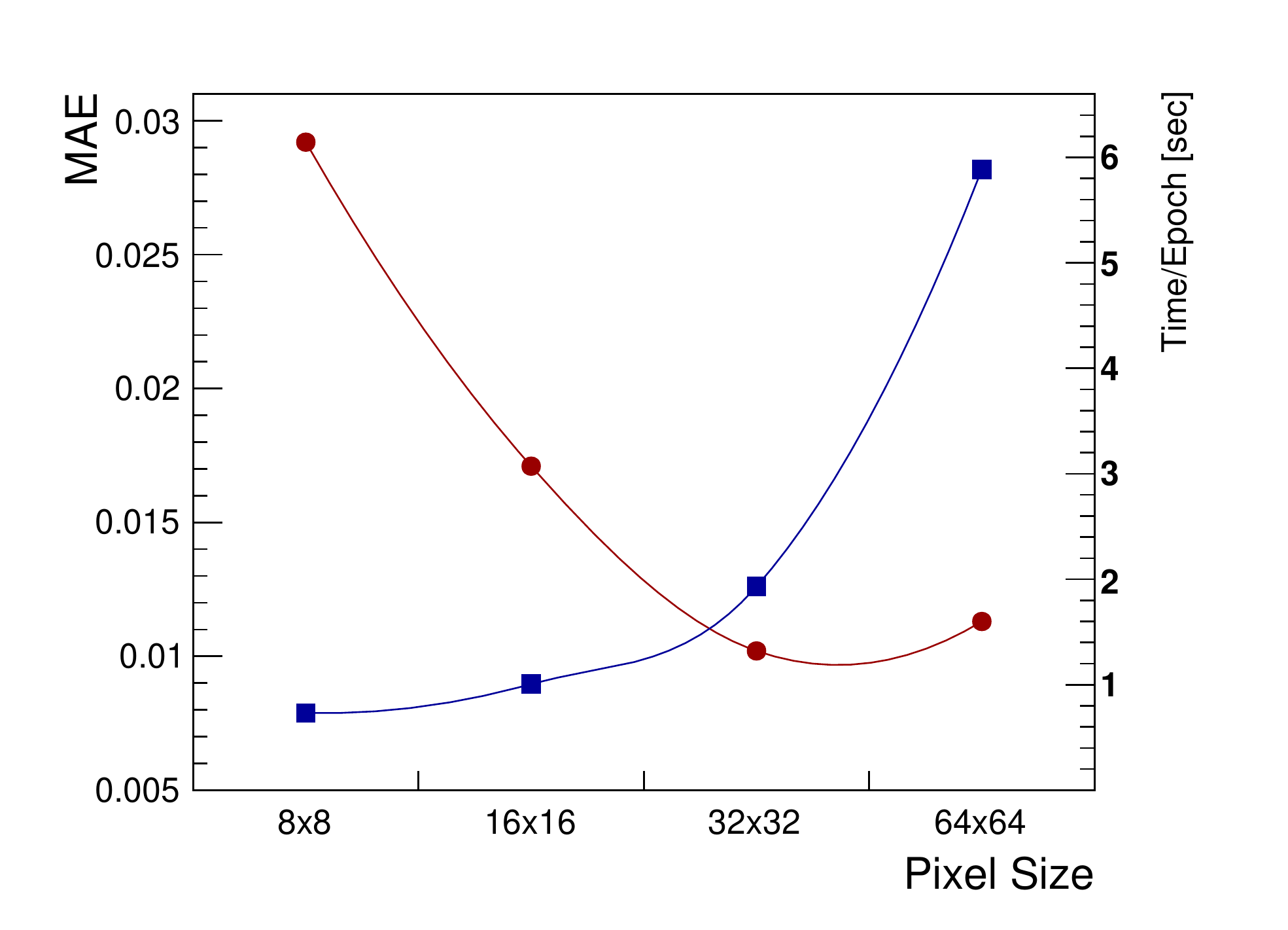}
\caption{(Color online) Mean absolute error (MAE) for $v_2$ (dots) and time taken per epoch (squares) during the training of the DNN as a function of input pixel size.}
\label{fig3}
\end{figure}

\begin{figure*}[ht!]
\includegraphics[scale=0.7]{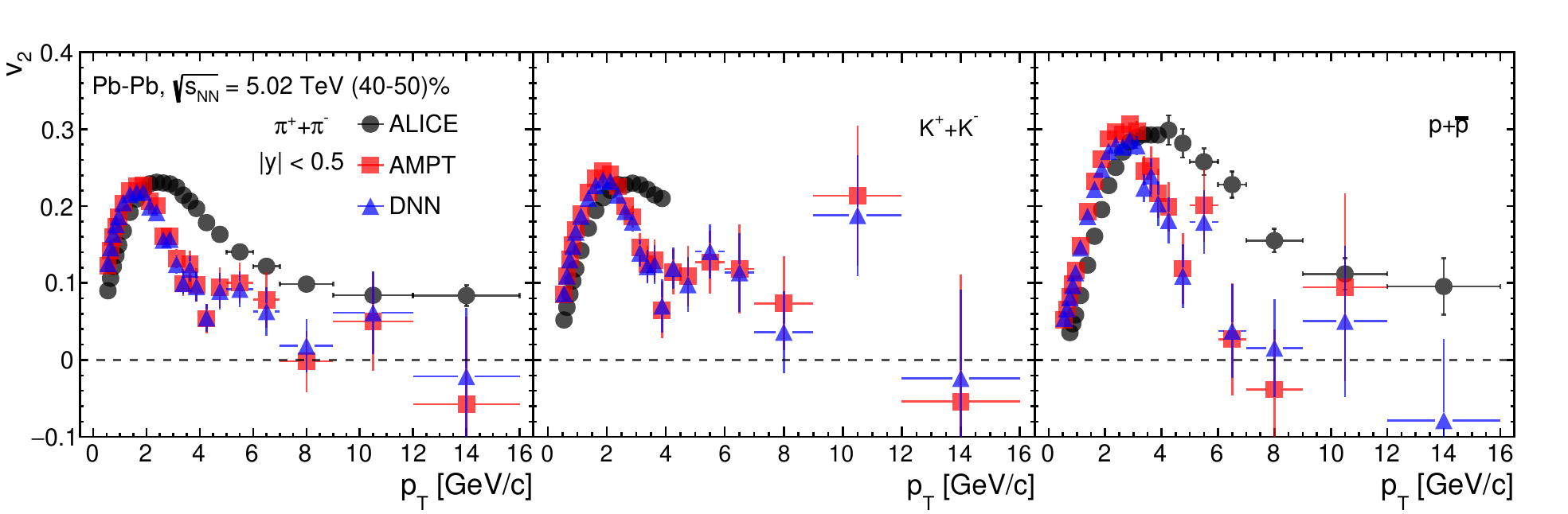}
\caption{(Color online) Elliptic flow ($v_2$) as a function of transverse momentum ($p_{\rm T}$) for $\pi^{\pm}$, $\rm K^{\pm}$, and $\rm p+\bar{p}$ in (40-50)\% central Pb-Pb collisions at $\sqrt{s_{\rm NN}} = 5.02$ TeV. ALICE~\cite{ALICE:2018yph}, AMPT, and DNN results are shown as dots, squares and triangles, respectively.}
\label{fig4}
\end{figure*}

Figure~\ref{fig3} shows the choice of different bin sizes in $(\eta-\phi)$ space for the input with corresponding mean absolute error (MAE) and the estimation for training time as time/epoch in seconds. One can infer from the figure that as the bin size increases, MAE decreases and corresponding training time increases. However, the bin size $32\times 32$ gives the optimum MAE and training time, which is used as the input binning for $p_{\rm T}$, mass and $\log\sqrt{(s_{NN}/s_{0})}$ weighted in $(\eta-\phi)$ space. Thus, this results in a total of 3072 ($= 32 \times 32 \times 3$) features per event as the input features. It is to be noted that the training time may differ from Fig.~\ref{fig3}, depending upon the type of CPU used. Time per epoch shown in Fig.~\ref{fig3} employs an Intel(R) Core(TM) i5-8279U (released Q2’19) CPU having four cores (eight threads) clocked at base frequency 2.40 GHz and a max turbo boost frequency of 4.10 GHz \cite{Intelcorei5}. The system has 8 GB of LPDDR3 RAM clocked at 2133 MHz. The data set with 3072 features per event is normalized with the L2 norm before DNN is ready to take it as input. 

Further quality assurance checks on the DNN model can be found in Ref.~\cite{Mallick:2022alr}, which suggests that the model can successfully be used not only for different centralities but also in different energy data and gives reasonable results for $p_{\rm T}$ dependent $v_2$ while having trained with minimum bias events in Pb--Pb collisions at $\sqrt{s_{\rm NN}}$ = 5.02 TeV data, event-by-event.
 
In this study, for the implementation of the DNN model, we use the KERAS v2.7.0 deep learning Application Programming Interface (API)~\cite{keras} with TensorFlow v2.7.0~\cite{Abadi:2016kic} in PYTHON along with the Scikit-Learn ML framework~\cite{sklearn}.

\section{Results and Discussions}
\label{sec4}
In this section, we begin with the comparison of the elliptic flow of $\pi^{\pm}$, $\rm K^{\pm}$, and $\rm p+\bar{p}$ predicted from the DNN estimator with the true values from AMPT simulation and ALICE results. We proceed to show the centrality, energy, and transverse momentum dependence of $v_2$, and also study whether the scaling properties of $v_2$ are encoded in the neural network. The evolution of the crossing point in $p_{\rm T}$ for baryon-meson $v_2$ separation in intermediate-$p_{\rm T}$ and the effect of transverse momentum-dependent training are also also described in this section.

\subsection{Comparing DNN predictions to AMPT and experimental data}

Figure~\ref{fig4} shows the elliptic flow, $v_2(p_{\rm T})$ for identified $\pi^{\pm}$, $\rm K^{\pm}$, and $\rm p+\bar{p}$ in (40-50)\% central Pb-Pb collisions at $\sqrt{s_{\rm NN}} = 5.02$ TeV. To compare the AMPT and DNN outcomes with ALICE results~\cite{ALICE:2018yph}, tracks with $p_{\rm T}> 0.5$ GeV/c in midrapidity, $|y|<0.5$ are considered for this case. The magnitude of $v_2(p_{\rm T})$ increases with increasing $p_{\rm T}$ for all three particle species till it attains a maximum value around $p_{\rm T} \approx 2.0$ GeV/c, and then it starts to decrease beyond this point. The values of $v_2(p_{\rm T})$ from AMPT obtained in this region (\textit{i.e.} $p_{\rm T} \lesssim 2.0\rm{-}3.0$ GeV/c) is comparable in magnitude with ALICE results for the individual particle cases. However, beyond this transverse momentum value, 
AMPT fails to describe the data as $v_2(p_{\rm T})$ falls faster with increasing $p_{\rm T}$ since fragmentation becomes the preferred mode of hadronization at high-$p_{\rm T}$.

It is interesting to note that DNN predictions agree with AMPT values quite nicely up to $p_{\rm T}\lesssim 4.0\rm{-}6.0$ GeV/c. Beyond this $p_{\rm T}$, the values from DNN start to differ from the AMPT values. The DNN model is trained with Pb-Pb collisions at $\sqrt{s_{\rm NN}} = 5.02$ TeV minimum bias events with the selection of all charged particles having $p_{\rm T}> 0.15$ GeV/c in pseudorapidity, $|\eta|<0.8$ simulated with AMPT. It is seen that the statistics in each $p_{\rm T}$-bin for all three particle types decreases to about a few 100 counts for $p_{\rm T} \gtrsim 6.0$ GeV/c, which presents a few instances to the DNN model during the training process. For this reason, the mismatch between DNN and AMPT comes into the picture for $p_{\rm T} \gtrsim 6.0$ GeV/c. For the above-mentioned reasons, this analysis is kept in the range $p_{\rm T} < 3.0$ GeV/c for the rest of the plots where the Monte Carlo and data are comparable.

\subsection{Centrality and energy dependence of elliptic flow}

\begin{figure*}[ht!]
\includegraphics[scale=0.7]{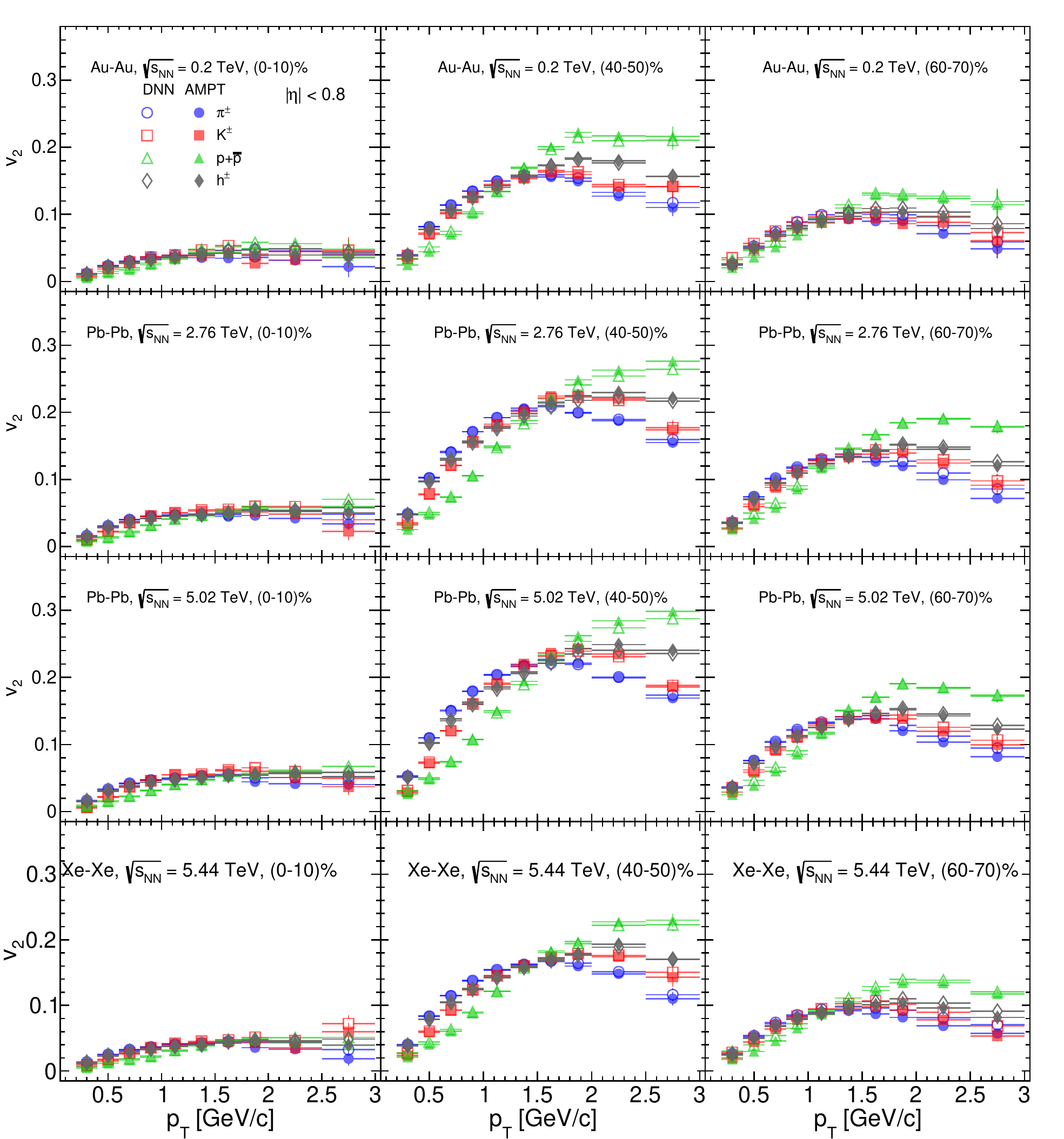}
\caption{(Color online) Centrality dependence of $v_2({p_{\rm T})}$ for $\pi^{\pm}$, $\rm K^{\pm}$, and $\rm p+\bar{p}$, and all charged hadrons ($h^{\pm}$). The values from AMPT and the predictions from DNN are shown.}
\label{fig5}
\end{figure*}

Figure~\ref{fig5} represents the centrality dependence of $v_2({p_{\rm T})}$ for $\pi^{\pm}$, $\rm K^{\pm}$, and $\rm p+\bar{p}$, and all charged hadrons ($h^{\pm}$) in Au-Au collisions at $\sqrt{s_{\rm NN}} = 200 $ GeV, Pb-Pb collisions at $\sqrt{s_{\rm NN}} = 2.76$, and $5.02$ TeV, and Xe-Xe collisions at $\sqrt{s_{\rm NN}} = 5.44$ TeV for (0-10)\%, (40-50)\%, and (60-70)\% centrality classes. The true values from AMPT and the predicted values from the DNN estimator are presented. As one moves from central to peripheral collisions, the magnitude of $v_2({p_{\rm T})}$ keeps on increasing for all particle types up to mid-central collisions, \textit{i.e.} (40-50)\%, and then it starts to decrease. This is well understood from the fact that initial geometrical anisotropy in the nuclear overlap region keeps increasing for peripheral collisions. However, for more peripheral collisions, \textit{i.e.} (60-70)\%,  $v_2({p_{\rm T})}$ gets reduced as the smaller size and shorter lifetime of the system do not allow the generation of large $v_2$. For $p_{\rm T} \lesssim 1.5$ GeV/c, there is a mass ordering of $v_2({p_{\rm T})}$ for different particles, meaning lighter particles have more $v_2$ than the heavier ones, here 
\begin{equation}
v_2^{\pi^{\pm}}>v_2^{K^{\pm}}>v_2^{p+\bar{p}}.
\end{equation}
This is understood as an effect of strong radial flow, which imposes an azimuthally symmetric velocity boost to all particles along with the anisotropic flow in the medium. In the intermediate-$p_{\rm T}$, baryon-meson $v_2$ gets separated as
\begin{equation}
v_2^{\rm(Baryons)} > v_2^{\rm(Mesons)}.
\end{equation}
 The baryon-meson $v_2$ separation could arise from the constituent quark coalescence mechanism embedded in the AMPT string melting mode~\cite{Tian:2009wg}. The DNN estimator, trained at the LHC energy, in Pb-Pb collisions at $\sqrt{s_{\rm NN}} = 5.02$ TeV minimum bias data set, can predict the identified particle $v_2$ for different collision systems at different collision energies. The mass ordering at low-$p_{\rm T}$ and the grouping of particles based on their constituent quarks in the intermediate-$p_{\rm T}$ is learned and preserved by the DNN estimator. 

\begin{figure*}[ht!]
\includegraphics[scale=0.7]{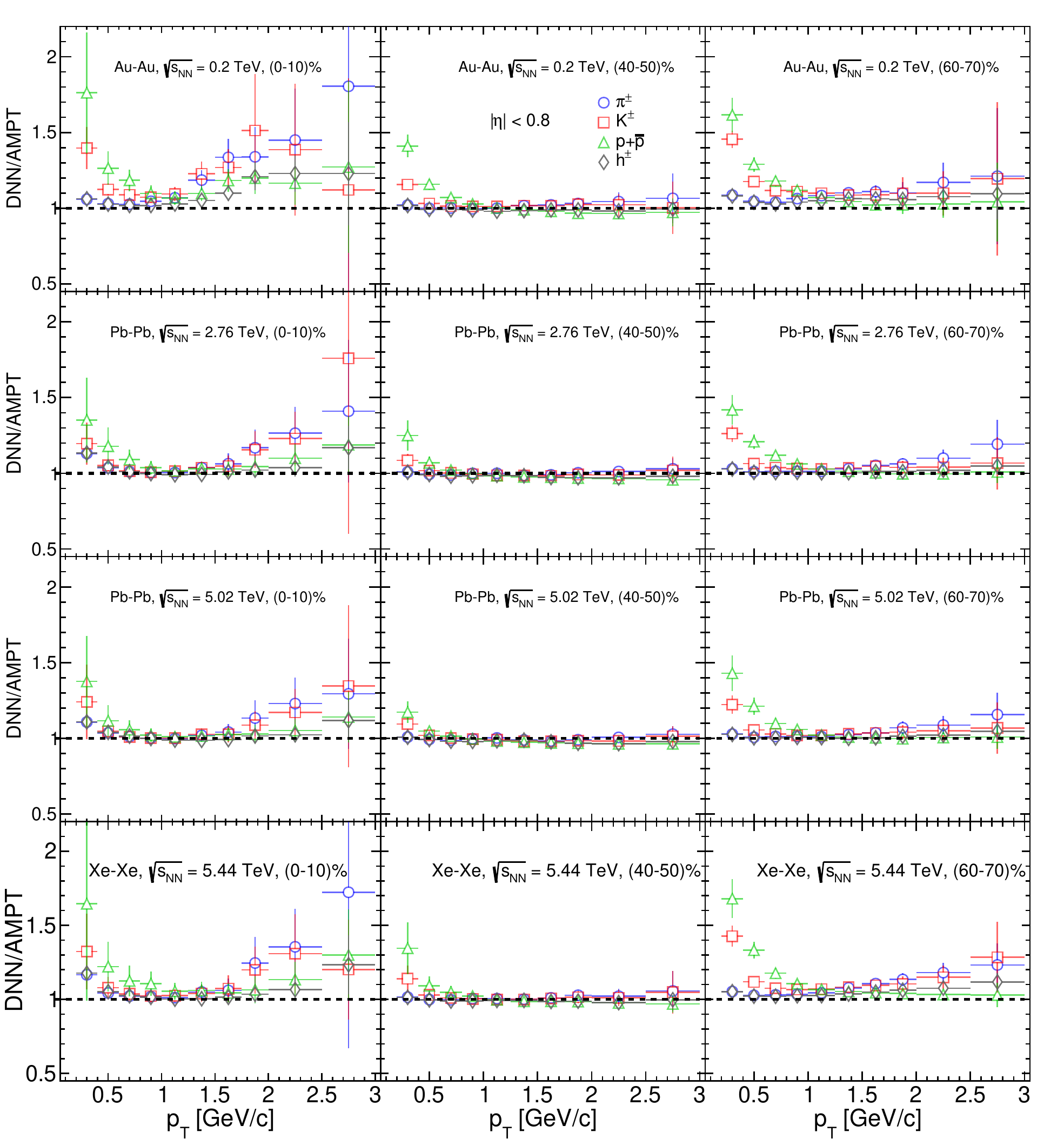}
\caption{(Color online) DNN to AMPT ratio of $v_2({p_{\rm T})}$ for $\pi^{\pm}$, $\rm K^{\pm}$, and $\rm p+\bar{p}$, and all charged hadrons ($h^{\pm}$).}
\label{fig6}
\end{figure*}

Figure~\ref{fig6} shows the ratio of $v_2(p_{\rm T})$ estimated with the DNN model over the AMPT values for $\pi^{\pm}$, $\rm K^{\pm}$, and $\rm p+\bar{p}$, and all charged hadrons ($h^{\pm}$). For the (40-50)\% centrality class, the DNN predictions for all particle types are in better agreement with the AMPT values as compared to (0-10)\% and (60-70)\% centrality classes. In Ref.~\cite{Mallick:2022alr}, while dealing with the centrality dependence of $v_2$ for all charged hadrons, one can observe that the predictions for the most central and peripheral cases have larger statistical uncertainty than the mid-central cases. Here, Fig.~\ref{fig6} also suggests similar behavior. We argue that this is entirely statistics driven as the number of events in the extreme centrality bins is less in a minimum bias data set; hence, with the minimum bias training, the DNN model does not get enough examples to learn from these extreme domains. A similar argument could also be given for the slight mismatch of proton $v_2$ at low-$p_{\rm T}$ and pion $v_2$ at intermediate-$p_{\rm T}$, where the respective statistics is comparably less. The solution to this problem could come from various collaborative learning models~\cite{collabgsong2018}. The same DNN model could be trained separately with centrality-wise data sets, unlike the minimum bias training to obtain a common set of model parameters. This could possibly take out the training sample bias due to the different number of events in different centrality bins, leading to an unequal level of information passing to the DNN training in different centrality bins.

\subsection{Constituent quark number scaling}

\begin{figure*}[ht!]
\includegraphics[scale=0.7]{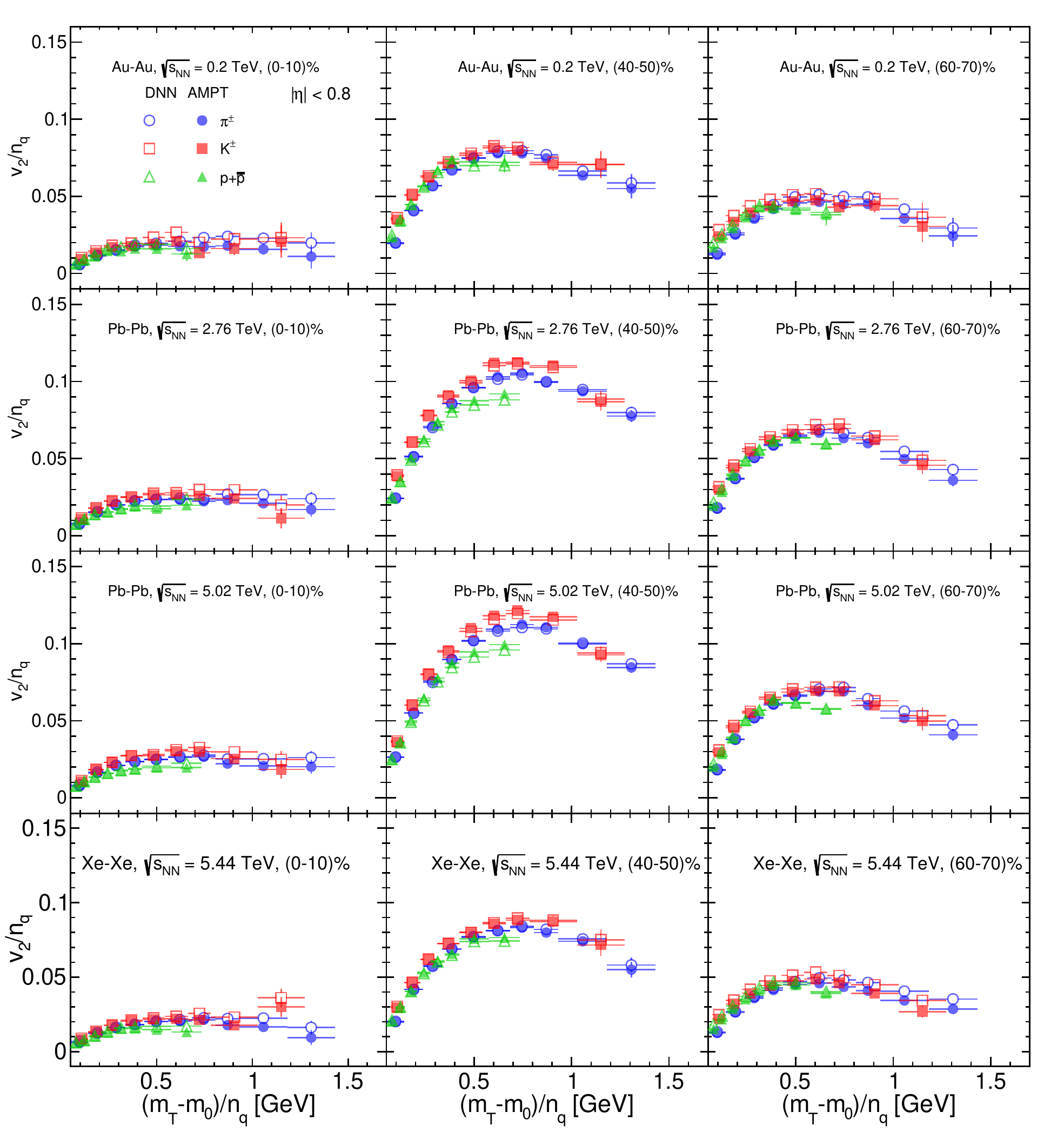}
\caption{(Color online) Centrality dependence of constituent quark scaling of $v_2$ as a function of $KE_{\rm T}/n_q$. The results from AMPT and the corresponding predictions from DNN are shown.}
\label{fig7}
\end{figure*}

The study of centrality and $p_{\rm T}$ dependence of $v_2$ for various identified particles seems incomplete without exploring its constituent quark number (NCQ) scaling behavior. By recalling our understanding of the specific mass ordering of $v_2(p_{\rm T})$ at lower $p_{\rm T}$ regime, if the mass ordering is driven by a hydrodynamic pressure gradient, then $v_2(KE_{\rm T})$ at lower-$p_{\rm T}$ should scale to a common set of values for all the particles, which is indeed observed in experiments~\cite{PHENIX:2006dpn}. Here, $KE_{\rm T}(= m_{\rm T} - m_{0})$ is the transverse kinetic energy, and $m_{\rm T}(=\sqrt{p_{\rm T}^2+m_{0}^2})$ and $m_{0}$ are the transverse mass and rest mass of the particle, respectively. At the intermediate-$p_{\rm T}$ regime, $v_2(KE_{\rm T})$ splits into two branches separately grouped by baryons and mesons. In this domain, the constituent quarks of the particles play a more substantial role in generating $v_2$ in hadrons than its mass. 
When the NCQ-scaling is applied to $v_2(KE_{\rm T})$, it shows a better scaling behaviour than $v_2(p_{\rm T})$ at lower-$p_{\rm T}$~\cite{PHENIX:2006dpn,PHENIX:2012swz}. Figure~\ref{fig7} shows the centrality dependence of $v_2/n_q$ as a function of $KE_{\rm T}/n_q$. Here, the number of constituent quarks, $n_q=2$ for mesons and $n_q=3$ for baryons. One can see that, at lower-$KE_{\rm T}/n_q$, the scaling is valid for all the particle types. However, the proton seems to break the scaling at higher-$KE_{\rm T}/n_q$. The violation of $KE_{\rm T}$ scaling is observed in experiments, so the results from AMPT are in line~\cite{PHENIX:2012swz,Mallick:2021hcs}. Here, it is interesting to note that the DNN can learn particle species-dependent scaling behavior as it closely agrees with the AMPT values for all energy and centrality classes. 

\subsection{Evolution of the crossing point}

\begin{figure}[ht!]
\includegraphics[scale=0.35]{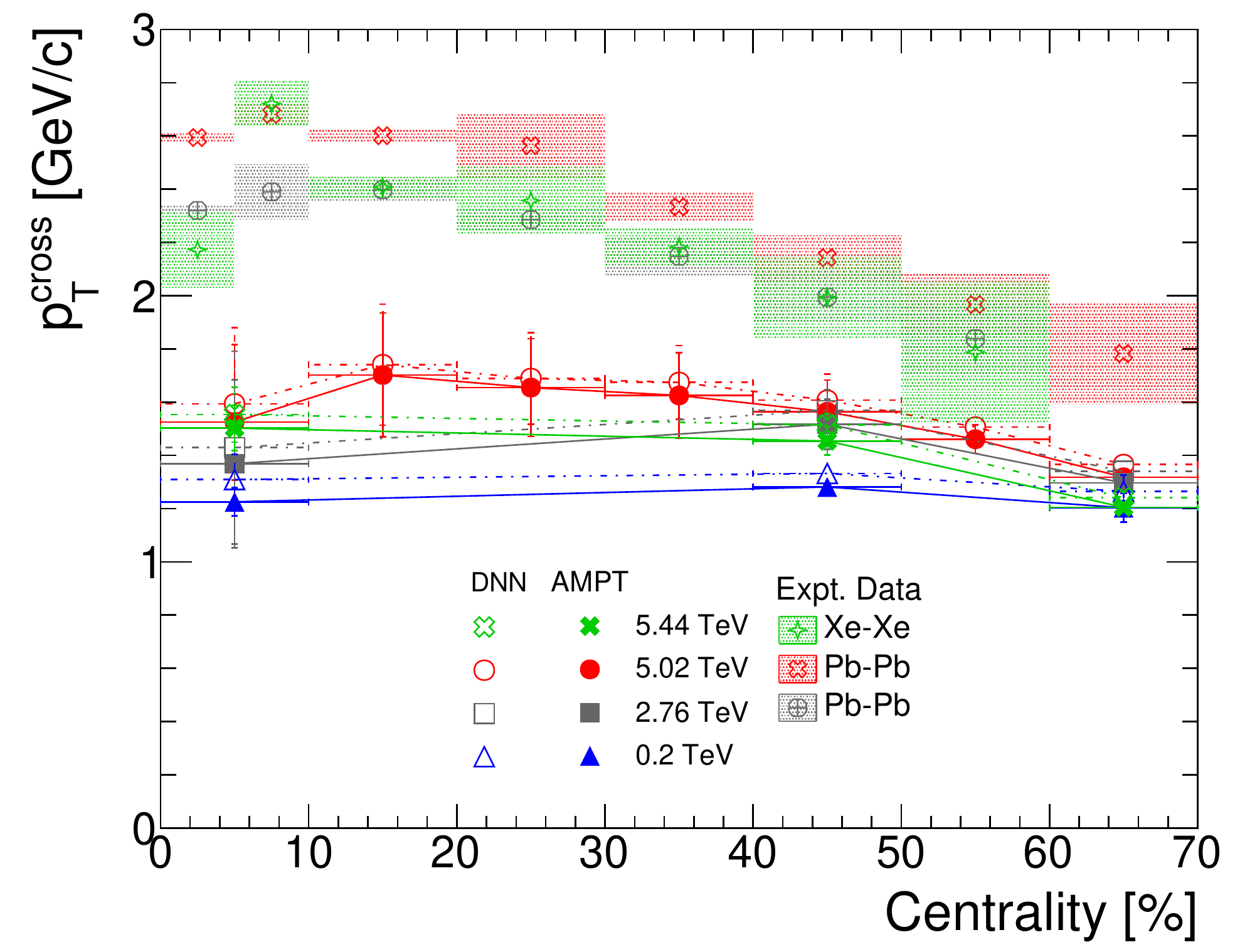}
\caption{(Color online) Centrality and energy dependent evolution of $p_{\rm T}-$crossing point, ($p_{\rm T}^{\rm cross}$) of baryon-meson elliptic flow in the intermediate-$p_{\rm T}$ regime. Here, the experimental data are obtained from the ALICE results~\cite{ALICE:2014wao,ALICE:2018yph,ALICE:2021ibz}.}
\label{fig8}
\end{figure}

The separation of baryon-meson elliptic flow at the intermediate-$p_{\rm T}$ is usually attributed to the quark coalescence/recombination mechanism of hadronization. It suggests an apparent dependence of hadron flow on the number of constituent quarks of the hadron. This leads to a higher flow of baryons than mesons and a relative enhancement of baryon yield over meson yield in the intermediate-$p_{\rm T}$ regime. Now, by looking into the crossing point in $p_{\rm T}$ to know precisely where the separation of baryon-meson flow occurs, one can confirm the coalescence picture if the crossing occurs at somewhat higher transverse momenta for higher centralities. Figure~\ref{fig8} shows the centrality dependence of the $p_{\rm T}$-crossing point of pion and proton elliptic flow for various collision systems at various energies. The vertical error bars represent the difference between the pion-proton and kaon-proton $p_{\rm T}$-crossing points. One can observe that the crossing point gradually shifts to higher-$p_{\rm T}$ as one move from peripheral to central collisions for all collision energies. The shift of the curve to higher-$p_{\rm T}$ with increasing energy could hint towards producing a denser partonic medium and the effect of increased radial flow with increasing collision energy. Here, the AMPT model qualitatively reproduces the trend of $p_{\rm T}$-crossing curve as seen in the ALICE results. A similar observation is made from the DNN predictions, which also closely match the AMPT curves for the respective collision systems and energies. The DNN model not only learns the splitting of the elliptic flow into two branches grouped separately by baryons and mesons but also preserves the same $p_{\rm T}$-crossing point of the separation, thus adding a more quantitative hold on the estimation of elliptic flow for identified charged hadrons.   

\begin{figure*}[ht!]
\includegraphics[scale=0.7]{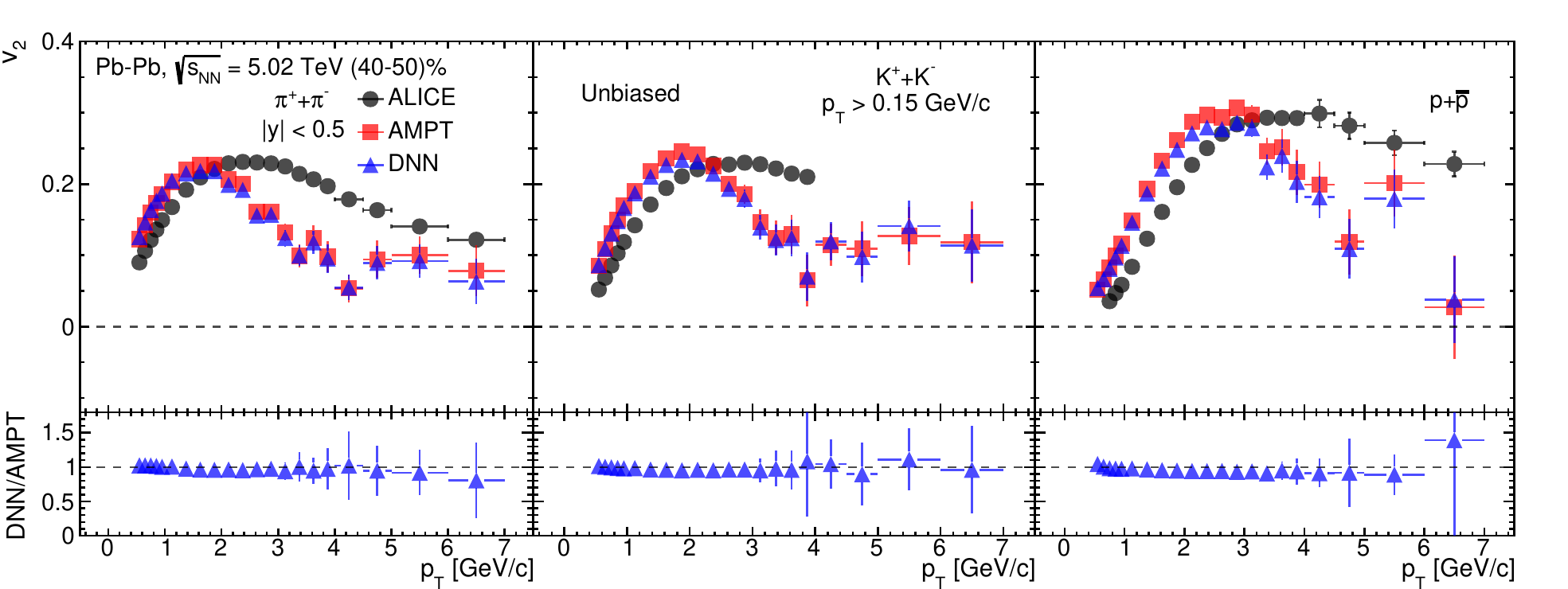}
\includegraphics[scale=0.7]{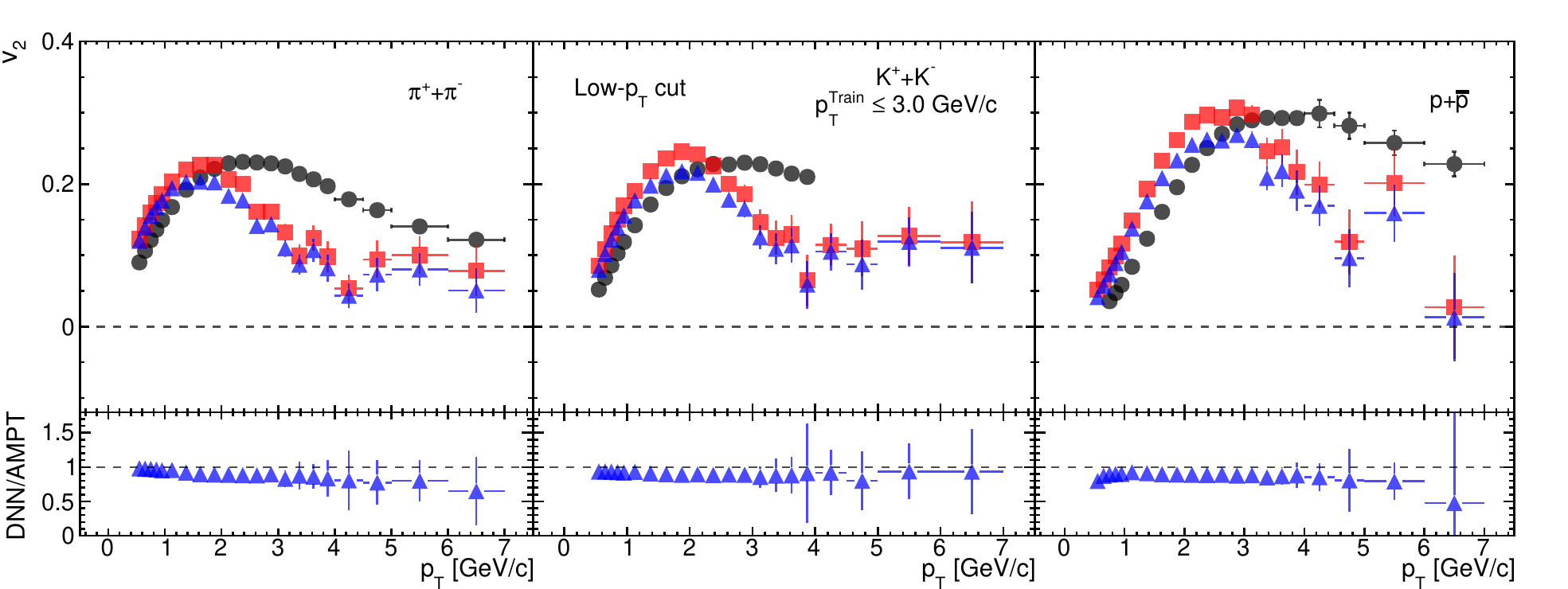}
\includegraphics[scale=0.7]{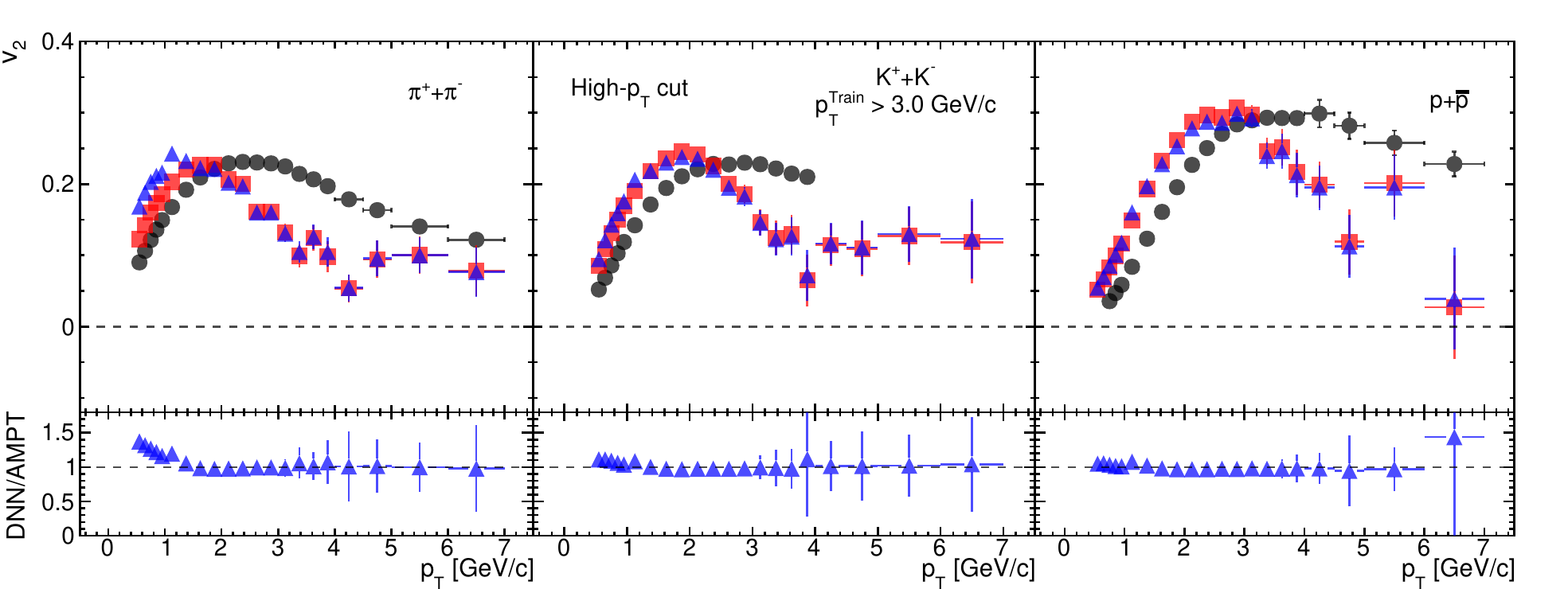}
\caption{(Color online) Evaluation of DNN with transverse momentum dependent training applied to (40-50)\% central Pb-Pb collisions at $\sqrt{s_{\rm NN}} = 5.02$ TeV. Results for the models with unbiased ($p_{\rm T}^{\rm Train} > 0.15$ GeV/c), low-$p_{\rm T}$ cut group ($p_{\rm T}^{\rm Train} \leq 3.0$ GeV/c), and high-$p_{\rm T}$ cut group ($p_{\rm T}^{\rm Train} > 3.0$ GeV/c) particles are shown.} 
\label{fig9}
\end{figure*}

\subsection{Effect of transverse momentum dependent training on DNN}
In Figure~\ref{fig9}, we present the predictions for elliptic flow as a function of transverse momentum from three different DNN models, each trained with a different set of inputs based on specific $p_{\rm T}$-cuts. These models are trained on the minimum bias Pb-Pb collisions at $\sqrt{s_{\rm NN}} = 5.02$ TeV. For this investigation, the same model architecture with similar hyperparameter settings is used for training. The only difference would be the obtained weights and biases of the network for different $p_{\rm T}$-cuts. 

The first row of Figure~\ref{fig9} refers to the predictions from the DNN model trained with particle kinematic information for tracks in $|\eta|<0.8$ with $p_{\rm T} > 0.15$ GeV/c. We name it the unbiased case. It shows the evolution of $v_{2}(p_{\rm T})$ in (40-50)\% central Pb-Pb collisions at $\sqrt{s_{\rm NN}} = 5.02$ TeV, where the value of $v_2$ has a maxima around $2.0 \lesssim p_{\rm T}|_{v_{2}^{\rm max}} \lesssim 3.0$ GeV/c depending on the particle species. As the training is done for an unbiased range of $p_{\rm T}$ which also includes this $p_{\rm T}|_{v_{2}^{\rm max}}$, the DNN model seems to capture the overall trend of the $v_{2}(p_{\rm T})$ including the transition behavior around $p_{\rm T}|_{v_{2}^{\rm max}}$. Thus the ratio of DNN to AMPT remains fairly close to unity for the entire $p_{\rm T}$-range (shown up to 7 GeV/c) for all the particle species. So far, all the above-discussed results are obtained from the DNN model with this unbiased training.  
 
Now, we study the effect of training the DNN model in different transverse momentum regions. The aim is to obtain two different DNN models trained with and without the transition point of $p_{\rm T}|_{v_{2}^{\rm max}}$ and then test their applicability to the estimation of $v_2(p_{\rm T})$ curve for the full range of $p_{\rm T}$. We can obtain two groups of particles in an event based on the transverse momentum cuts applied to the tracks and we name them as low-$p_{\rm T}$ cut group ($p_{\rm T}^{\rm Train}\leq3.0$ GeV/c) and high-$p_{\rm T}$ cut group ($p_{\rm T}^{\rm Train}>3.0$ GeV/c). The estimation of elliptic flow is also performed using only these selected particles for the respective groups. The low-$p_{\rm T}$ cut group includes the $p_{\rm T}|_{v_{2}^{\rm max}}$, thus, the transition behaviour of $v_2(p_{\rm T})$ around $p_{\rm T}|_{v_{2}^{\rm max}}$. However, the high-$p_{\rm T}$ cut group does not include this transition behavior directly from particle kinematics. After the training of these two separate DNN models with low-$p_{\rm T}$ and high-$p_{\rm T}$ particles from Pb-Pb collisions at $\sqrt{s_{\rm NN}} = 5.02$ TeV minimum bias events, we apply each model to predict the entire $v_2(p_{\rm T})$ curve for the full range of $p_{\rm T}$ (shown up to 7 GeV/c) for the (40-50)\% central Pb-Pb collisions at $\sqrt{s_{\rm NN}} = 5.02$ TeV. 
 
From the second and third row of Figure~\ref{fig9}, it is quite interesting to see that both the models, although being trained with two disjoint sets of particle groups, can successfully reproduce the entire $v_2(p_{\rm T})$ curve up to a reasonable extent in their respective cases. Somehow the models learn and complement the missing information of the $v_2(p_{\rm T})$ curve. For the low-$p_{\rm T}$ cut group, the missing factor in training is the high-$p_{\rm T}$ particles, and for the high-$p_{\rm T}$ cut group, it is the low-$p_{\rm T}$ particles, and also the absence of the $p_{\rm T}|_{v_{2}^{\rm max}}$. As the information encoded in the network behaves like a generalized global curve of $v_2$ applicable to the entire $p_{\rm T}$ domain, the DNN models can extrapolate any regions of interest regardless of their training. This global curve learned by the DNN models represents the dependence of $v_2$ on transverse momentum and also its transition behavior around the $p_{\rm T}|_{v_{2}^{\rm max}}$. Although the mathematical function for such a curve is not available in the literature, the DNN can map this function from the existing input-output correlations irrespective of its training domain.

Another observation is the behavior of elliptic flow for different particle species. When the DNN model trained with low-$p_{\rm T}$ particles is applied to predict $v_2$ for pion and kaon, a slight deviation is observed between DNN and AMPT starting from intermediate to high-$p_{\rm T}$. This is visible in the bottom ratio plots in the second row of Fig.~\ref{fig9}. It hints that the elliptic flow for pion and kaon primarily originates from low- to intermediate-$p_{\rm T}$. However, in the same figure, the prediction for proton deviates even at very low-$p_{\rm T}$, although the model includes this low-$p_{\rm T}$ region in training. It suggests that the contribution to elliptic flow for proton mainly originates from high-$p_{\rm T}$. Thus by removing this region in the low-$p_{\rm T}$ cut DNN model, the accuracy is reduced for the entire curve for proton $v_2$.

This hypothesis can be verified with the DNN model trained with high-$p_{\rm T}$ particles. One can see from the third row of Fig.~\ref{fig9} that the prediction of elliptic flow for proton is now more accurate for the entire $p_{\rm T}$ range as this DNN model includes high-$p_{\rm T}$ particles. This observation supports our earlier statement of DNN captures a global curve of $v_2(p_{\rm T})$ and proton $v_2$ has a dominant contribution from high-$p_{\rm T}$. However, in the same figure, pion and kaon elliptic flow deviate more in the low-$p_{\rm T}$ as the necessary contribution to elliptic flow is absent in this region for the high-$p_{\rm T}$ cut model, and same is visible in the ratio plots.

\section{Summary}
\label{sec5}

To sum everything up that has been stated so far, this work demonstrates the applicability and accuracy of a DNN-based machine learning model to evaluate the second-order anisotropic flow coefficient ($v_2$) for identified particles from final state particle kinematic information in heavy-ion collisions. The DNN model can eminently estimate $v_2$ for light-flavor identified particles such as $\pi^{\pm}$, $\rm K^{\pm}$, and $\rm p+\bar{p}$ in heavy-ion collisions at RHIC and LHC energies. The prediction accuracy of the DNN model is not only limited to Pb-Pb collisions at $\sqrt{s_{\rm NN}}$ = 5.02 TeV at which the model is trained but also applicable to other systems such as Pb-Pb collisions at $\sqrt{s_{\rm NN}}$ = 2.76 TeV, Xe-Xe collisions at $\sqrt{s_{\rm NN}}$ = 5.44 TeV, and Au-Au collisions at $\sqrt{s_{\rm NN}}$ = 200 GeV. From the minimum bias training, the model could successfully learn and preserve the centrality, energy, and transverse momentum dependence of elliptic flow for other collision systems at various energies. This becomes a striking feature of DNN-based machine learning, where, while dealing with multihadron production dynamics, the 
machine is trained with the control parameters like the collision energy, impact parameter (here, the number of participants or the final state charged particle multiplicity), particle mass, and the transverse momentum. 
In addition, the DNN model retains the constituent quarks number scaling behavior for elliptic flow as a function of transverse kinetic energy for the respective collision systems and energies. The DNN model also quantitatively preserves the $p_{\rm T}$-crossing point of baryon-meson $v_2$ curves at intermediate-$p_{\rm T}$, which scales with centrality and center-of-mass energy of collisions for various systems. Finally, by training the DNN models with different kinematic regions based on specific $p_{\rm T}$-cuts, and then applying it to the full $p_{\rm T}$ range, we arrive at the conclusion that there exists a global functional dependency of elliptic flow with transverse momentum, as the DNN model trained in a certain kinematic region could describe the overall curve for the full kinematic domain. We also learn that the elliptic flow of pion and kaon originates mainly from the low-$p_{\rm T}$ region, whereas the elliptic flow of the proton is dominated due to the high-$p_{\rm T}$ region. 



\section*{Acknowledgements}
S.~P. acknowledges the doctoral fellowship from UGC, Government of India. R.~S. sincerely acknowledges the DAE-DST, Government of India funding under the mega-science project – “Indian participation in the ALICE experiment at CERN” bearing Project No. SR/MF/PS-02/2021-IITI (E-37123). G.~G.~B. gratefully acknowledges the Hungarian National Research, Development and Innovation Office (NKFIH) under Contracts No. OTKA K135515, No. NKFIH 2019-2.1.11-TET-2019-00078, No. 2019-2.1.11-TET-2019-00050, and NEMZ\_KI-2022-00009 and Wigner Scientific Computing Laboratory (WSCLAB, the former Wigner GPU Laboratory). The authors gratefully acknowledge the Memorandum of understanding (MoU) between IIT Indore and Wigner Research Centre for Physics (WRCP), Hungary, for the techno-scientific international cooperation.


\begin{thebibliography}{100}

\bibitem{Bass:1998vz}
S.~A.~Bass, M.~Gyulassy, H.~Stoecker and W.~Greiner,
J. Phys. G \textbf{25}, R1-R57 (1999). 

\bibitem{Heinz:2013th}
U.~Heinz and R.~Snellings,
Ann. Rev. Nucl. Part. Sci. \textbf{63}, 123 (2013).

\bibitem{STAR:2003wqp}
J.~Adams \textit{et al.} [STAR Collaboration],
Phys. Rev. Lett. \textbf{92}, 052302 (2004).

\bibitem{ALICE:2010suc}
K.~Aamodt \textit{et al.} [ALICE Collaboration],
Phys. Rev. Lett. \textbf{105}, 252302 (2010).

\bibitem{ALICE:2011ab}
K.~Aamodt \textit{et al.} [ALICE Collaboration],
Phys. Rev. Lett. \textbf{107}, 032301 (2011).

\bibitem{ALICE:2014dwt}
B.~B.~Abelev \textit{et al.} [ALICE Collaboration],
Phys. Rev. C \textbf{90}, 054901 (2014).

\bibitem{Kolb:2000fha}
P.~F.~Kolb, P.~Huovinen, U.~W.~Heinz and H.~Heiselberg,
Phys. Lett. B \textbf{500}, 232 (2001).

\bibitem{Bass:2000ib}
S.~A.~Bass and A.~Dumitru,
Phys. Rev. C \textbf{61}, 064909 (2000).

\bibitem{Nonaka:2006yn}
C.~Nonaka and S.~A.~Bass,
Phys. Rev. C \textbf{75}, 014902 (2007).

\bibitem{Teaney:2001av}
D.~Teaney, J.~Lauret and E.~V.~Shuryak,
[arXiv:nucl-th/0110037 [nucl-th]].

\bibitem{Hirano:2005xf}
T.~Hirano, U.~W.~Heinz, D.~Kharzeev, R.~Lacey and Y.~Nara,
Phys. Lett. B \textbf{636}, 299 (2006).

\bibitem{Voloshin:2002wa}
S.~A.~Voloshin,
Nucl. Phys. A \textbf{715}, 379 (2003). 

\bibitem{Molnar:2003ff}
D.~Molnar and S.~A.~Voloshin,
Phys. Rev. Lett. \textbf{91}, 092301 (2003). 

\bibitem{Sato:1981ez}
H.~Sato and K.~Yazaki,
Phys. Lett. B \textbf{98}, 153 (1981).

\bibitem{Dover:1991zn}
C.~B.~Dover, U.~W.~Heinz, E.~Schnedermann and J.~Zimanyi,
Phys. Rev. C \textbf{44}, 1636 (1991).

\bibitem{PHENIX:2012swz}
A.~Adare \textit{et al.} [PHENIX Collaboration],
Phys. Rev. C \textbf{85}, 064914 (2012). 

\bibitem{ALICE:2014wao}
B.~B.~Abelev \textit{et al.} [ALICE Collaboration],
JHEP \textbf{06}, 190 (2015).


\bibitem{ALICE:2018yph}
S.~Acharya \textit{et al.} [ALICE Collaboration],
JHEP \textbf{09}, 006 (2018).

\bibitem{Poskanzer:1998yz}
A.~M.~Poskanzer and S.~A.~Voloshin,
Phys. Rev. C \textbf{58}, 1671 (1998).

\bibitem{Borghini:2000sa}
N.~Borghini, P.~M.~Dinh and J.~Y.~Ollitrault,
Phys. Rev. C \textbf{63}, 054906 (2001).

\bibitem{Bhalerao:2003xf}
R.~S.~Bhalerao, N.~Borghini and J.~Y.~Ollitrault,
Nucl. Phys. A \textbf{727}, 373 (2003).

\bibitem{Bhalerao:2014mua}
R.~S.~Bhalerao, J.~Y.~Ollitrault, S.~Pal and D.~Teaney,
Phys. Rev. Lett. \textbf{114}, 152301 (2015).

\bibitem{Hippert:2019swu}
M.~Hippert, D.~Dobrigkeit Chinellato, M.~Luzum, J.~Noronha, T.~Nunes da Silva and J.~Takahashi,
Phys. Rev. C \textbf{101}, 034903 (2020).

\bibitem{CMS:2017mzx}
A.~M.~Sirunyan \textit{et al.} [CMS Collaboration],
Phys. Rev. C \textbf{96}, 064902 (2017).

\bibitem{Gardim:2019iah}
F.~G.~Gardim, F.~Grassi, P.~Ishida, M.~Luzum and J.~Y.~Ollitrault,
Phys. Rev. C \textbf{100}, 054905 (2019).


\bibitem{Mallick:2022alr}
N.~Mallick, S.~Prasad, A.~N.~Mishra, R.~Sahoo and G.~G.~Barnaf\"oldi,
Phys. Rev. D \textbf{105}, 114022 (2022).

\bibitem{Lin:2004en}
Z.~W.~Lin, C.~M.~Ko, B.~A.~Li, B.~Zhang and S.~Pal,
Phys. Rev. C \textbf{72}, 064901 (2005).

\bibitem{ampthijing}
X.~N.~Wang and M.~Gyulassy,
Phys.\ Rev.\ D {\bf 44}, 3501 (1991).

\bibitem{amptzpc}
B.~Zhang, \ Comput. \ Phys. \ Commun. {\bf 109}, 193 (1998).

\bibitem{Lin:2001zk}
Z.~w.~Lin and C.~M.~Ko,
Phys. Rev. C \textbf{65}, 034904 (2002).

\bibitem{He:2017tla}
Y.~He and Z.~W.~Lin,
Phys. Rev. C \textbf{96}, 014910 (2017).

\bibitem{amptart1}
B.~Li, A.~T.~Sustich, B.~Zhang and C.~M.~Ko,
Int.\ J.\ Mod.\ Phys.\ E {\bf 10}, 267 (2001).

\bibitem{amptart2}
B.~A.~Li and C.~M.~Ko,
Phys.\ Rev.\ C {\bf 52}, 2037 (1995).

\bibitem{ampthadron2}
R.~J.~Fries, B.~Muller, C.~Nonaka and S.~A.~Bass,
Phys.\ Rev.\ Lett.\  {\bf 90}, 202303 (2003).

\bibitem{ampthadron3}
R.~J.~Fries, B.~Muller, C.~Nonaka and S.~A.~Bass,
Phys.\ Rev.\ C {\bf 68}, 044902 (2003).

\bibitem{Greco:2003mm}
V.~Greco, C.~M.~Ko and P.~Levai,
Phys. Rev. C \textbf{68}, 034904 (2003).

\bibitem{Tripathy:2018bib}
S.~Tripathy, S.~De, M.~Younus and R.~Sahoo,
Phys. Rev. C \textbf{98}, 064904 (2018).

\bibitem{Mallick:2020ium}
N.~Mallick, R.~Sahoo, S.~Tripathy and A.~Ortiz,
J. Phys. G \textbf{48}, 045104 (2021).

\bibitem{Mallick:2021wop}
N.~Mallick, S.~Tripathy, A.~N.~Mishra, S.~Deb and R.~Sahoo,
Phys. Rev. D \textbf{103}, 094031 (2021).

\bibitem{Prasad:2022zbr}
S.~Prasad, N.~Mallick, S.~Tripathy and R.~Sahoo,
[arXiv:2207.12133 [hep-ph]].

\bibitem{Loizides:2017ack}
C.~Loizides, J.~Kamin and D.~d'Enterria,
Phys. Rev. C \textbf{97}, 054910 (2018)
[erratum: Phys. Rev. C \textbf{99}, 019901 (2019)].

\bibitem{Voloshin:1994mz}
S.~Voloshin and Y.~Zhang,
Z. Phys. C \textbf{70}, 665 (1996).

\bibitem{Masera:2009zz}
M.~Masera, G.~Ortona, M.~G.~Poghosyan and F.~Prino,
Phys. Rev. C \textbf{79}, 064909 (2009).

\bibitem{Bhat:2010zz}
P.~C.~Bhat,
Ann. Rev. Nucl. Part. Sci. \textbf{61}, 281 (2011).

\bibitem{Roe:2004na}
B.~P.~Roe, H.~J.~Yang, J.~Zhu, Y.~Liu, I.~Stancu and G.~McGregor
Nucl. Instrum. Meth. A \textbf{543}, 577 (2005).

\bibitem{Baldi:2014kfa}
P.~Baldi, P.~Sadowski and D.~Whiteson,
Nature Commun. \textbf{5}, 4308 (2014).

\bibitem{Baldi:2014pta}
P.~Baldi, P.~Sadowski and D.~Whiteson,
Phys. Rev. Lett. \textbf{114}, 111801 (2015).

\bibitem{Buckley:2011kc}
A.~Buckley, A.~Shilton and M.~J.~White,
Comput. Phys. Commun. \textbf{183}, 960 (2012).

\bibitem{Bridges:2010de}
M.~Bridges, K.~Cranmer, F.~Feroz, M.~Hobson, R.~Ruiz de Austri and R.~Trotta,
JHEP \textbf{03}, 012 (2011).

\bibitem{biro}
G{\'a}bor~B{\'\i}r{\'o}, Mih{\'a}ly~Pocsai, Imre~F.~Barna, Gergely~G.~Barnaf{\"o}ldi, Joshua~T.~Moody, and G{\'a}bor Demeter, Optics \& Laser Technology \textbf{159}, 108948 (2023).

\bibitem{relu}
X.~Glorot, A.~Bordes, and Y.~Bengio, Proceedings of the Fourteenth International Conference on Artificial Intelligence and Statistics, Ft. Lauderdale, USA, PMLR Workshop and Conference Proceedings pp. 315–323 (2011). \url{https://proceedings.mlr.press/v15/glorot11a.html}

\bibitem{Kingma:2014vow}
D.~P.~Kingma and J.~Ba,
[arXiv:1412.6980 [cs.LG]].

\bibitem{Intelcorei5}
Available online: \url{https://www.intel.com/content/www/us/en/products/sku/191070/intel-core-i58279u-processor-6m-cache-up-to-4-10-ghz/specifications.html}

\bibitem{keras}
F.~Chollet, \textit{et al.} \url{https://keras.io/}

\bibitem{Abadi:2016kic}
M.~Abadi, A.~Agarwal, P.~Barham, E.~Brevdo, Z.~Chen, C.~Citro, G.~S.~Corrado, A.~Davis, J.~Dean and M.~Devin, \textit{et al.}
[arXiv:1603.04467 [cs.DC]]. \url{https://www.tensorflow.org/}

\bibitem{sklearn}
F.~Pedregosa \textit{et al.}, J.~Mach.~Learn.~Res. \textbf{12}, 2825 (2011). \url{https://dl.acm.org/doi/10.5555/1953048.2078195}

\bibitem{Tian:2009wg}
J.~Tian, J.~H.~Chen, Y.~G.~Ma, X.~Z.~Cai, F.~Jin, G.~L.~Ma, S.~Zhang and C.~Zhong,
Phys. Rev. C \textbf{79}, 067901 (2009).

\bibitem{collabgsong2018}
G.~Song, and W.~Chai, 
 [arXiv:1805.11761 [stat.ML]].

\bibitem{PHENIX:2006dpn}
A.~Adare \textit{et al.} [PHENIX Collaboration],
Phys. Rev. Lett. \textbf{98}, 162301 (2007).


\bibitem{Mallick:2021hcs}
N.~Mallick, S.~Tripathy and R.~Sahoo,
Eur. Phys. J. C \textbf{82}, 524 (2022).


\bibitem{ALICE:2021ibz}
S.~Acharya \textit{et al.} [ALICE Collaboration],
JHEP \textbf{10}, 152 (2021).

\end{thebibliography}
\end{document}